\documentclass[12pt]{article}

\usepackage{amssymb}
\usepackage{amsmath}
\usepackage{epsfig}
\usepackage{graphicx}
\usepackage{natbib}

\newfont{\bbold}{msbm10 scaled \magstep1}

\let\hat=\widehat

\newcommand{\dd}{{\rm d}}
\newcommand{\nn}{\nonumber}
\newcommand{\half}{\mbox{$\frac{1}{2}$}}

\title{{\bf Self-organised criticality in base-pair breathing
in DNA with a defect}}

\author{Ciprian-Ionu\c{t} DUDUIAL\u{A}, Jonathan A.D. WATTIS \\ 
School of Mathematical Sciences, University of Nottingham, \\ 
Nottingham, United Kingdom, NG7 2RD, UK \\ 
Charles A. LAUGHTON, \\ 
School of Pharmacy, University of Nottingham,\\ 
Nottingham, United Kingdom, NG7 2RD, UK}

\begin{document} 

\maketitle 

\begin{abstract}
We analyse base-pair breathing in a DNA sequence of 12 base-pairs
with a defective base at its centre. We use both all-atom molecular
dynamics (MD) simulations
and a system of stochastic differential equations (SDE).   In both cases,
Fourier analysis of the trajectories reveals self-organised critical
behaviour in the breathing of base-pairs.
The Fourier Transforms (FT) of the interbase distances show 
power-law behaviour with gradients close to $-1$.
The scale-invariant behaviour we have found provides evidence for the
view that base-pair breathing corresponds to the nucleation stage of
large-scale DNA opening (or 'melting') and that
this process is a (second-order) phase transition.
Although the random forces in our SDE system were introduced
as white noise, FTs of the displacements exhibit pink noise, as do the
displacements in the AMBER/MD simulations.
dynamics
MSC 60G18 Probability theory and stochastic processes: self-similar processes \\ 
MSC  37K40 Dynamical systems: Soliton theory, asymptotic behavior of solutions\\ 
MSC 92C45 Biology and other natural sciences: Kinetics in biochemical problems
\end{abstract}

\section{Introduction}

Much has been written about the process by which double-stranded 
DNA becomes two separated single strands, see, for example,  
\cite{hennig,amb}.  Several authors refer to the melting of DNA as  
having the form of a phase transition, 
in which the opening of one or a few base-pairs is akin to nucleation, 
and the subsequent growth of open `bubbles' is similar to the 
growth of crystal nucleii.   This two-stage process is 
relatively well understood in the contexts of crystal growth, and 
aerosol formation, but less so in kinetics of DNA replication. 
The aim of this paper is to show that the nucleation event -- that is the 
initial opening of bases -- exhibits self-similar, or scale-free, 
critical behaviour,  as one would expect at a phase transition. 
Whilst other studies have analysed bubble growth and the statistics 
of bubble length, 
we choose to focus on the temporal statistics.   Our model is more 
detailed, only focusing on 12 base-pairs and the opening of the first 
base; our model is thus significantly smaller than the bubble growth 
models of \cite{hennig,amb}; however, our models are more accurate 
in that the AMBER simulations \cite{Amber} (Assisted Model Building 
with Energy Refinement) include the effect of every atom in the 
DNA and the water molecules in the environment, and our SDE models 
include an accurate fitting of the nonlinear inter-base potential energy 
as described in our earlier work \cite{Duduiala1}.  

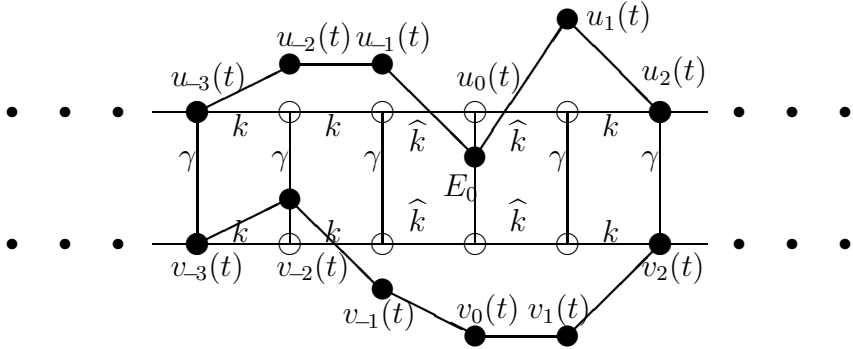
\begin{figure}[ht]
\begin{center}
\begin{picture}(520,105)(-60,-22)
\thinlines
\multiput(52,0)(35,0){6}{\circle{8}}
\multiput(52,50)(35,0){6}{\circle{8}}
\multiput(35,50)(35,0){6}{\line(1,0){35}}
\multiput(35,0)(35,0){6}{\line(1,0){35}}
\multiput(52,0)(35,0){6}{\line(0,1){50}}
\multiput( 45,30)(35,0){3}{$\gamma$}
\multiput(185,30)(35,0){2}{$\gamma$}
\put(145,18){$E_0$}
\multiput(65,41)(35,0){2}{$k$}
\multiput(205,41)(35,0){1}{$k$}
\multiput(205,1)(35,0){1}{$k$}
\multiput(65,1)(35,0){2}{$k$}
\put(170,35){$\hat k $}
\put(132,35){$\hat k$}
\put(170,5){$\hat k $}
\put(132,5){$\hat k$}
\thicklines
\put( 52, 50){\line(2,1){36}}
\put( 87, 68){\line(1,0){36}}
\put(122, 68){\line(1,-1){36}}
\put(157, 33){\line(2,3){36}}
\put(192, 85){\line(1,-1){36}}
\put(52,50){\circle*{8}}
\put(87,68){\circle*{8}}
\put(122,68){\circle*{8}}
\put(157,33){\circle*{8}}
\put(192,85){\circle*{8}}
\put(227,50){\circle*{8}}
\put( 42,60){$u_{\!-\!3}(t)$}
\put( 82,75){$u_{\!-\!2}(t)$}
\put(112,75){$u_{\!-\!1}(t)$}
\put(150,60){$u_0(t)$}
\put(199,83){$u_1(t)$}
\put(220,62){$u_2(t)$}
\put( 52,0){\line(2,1){36}}
\put( 87,17){\line(1,-1){36}}
\put(122,-17){\line(2,-1){36}}
\put(157,-35){\line(1,0){36}}
\put(192,-35){\line(1,1){36}}
\put( 42,-12){$v_{\!-\!3}(t)$}
\put(82,-12){$v_{\!-\!2}(t)$}
\put(107,-30){$v_{\!-\!1}(t)$}
\put(150,-28){$v_0(t)$}
\put(177,-28){$v_1(t)$}
\put(220,-12){$v_2(t)$}
\put(52,0){\circle*{8}}
\put(87,17){\circle*{8}}
\put(122,-17){\circle*{8}}
\put(157,-35){\circle*{8}}
\put(192,-35){\circle*{8}}
\put(227,0){\circle*{8}}
\multiput(257,0)(20,0){3}{\circle*{4}}
\multiput(257,50)(20,0){3}{\circle*{4}}
\multiput(22,0)(-20,0){3}{\circle*{4}}
\multiput(22,50)(-20,0){3}{\circle*{4}}
\end{picture}
\end{center}
\caption{IIllustration of the mesoscopic model of DNA.}
\label{ladder-fig}
\end{figure}

According to Watson $\&$ Crick \cite{Watson}, the structure of a 
DNA duplex consists of two chains of bases. These bases are of 
four types: the purines Adenine (A) and Guanine (G) and the 
pyrimidines Cytosine (C) and Thymine (T). Along the chains, 
the bases are linked by covalent bonds, while the opposite 
bases from the two chains pair together by two or three hydrogen 
bonds forming base-pairs. Only A-T or C-G pairs are possible. 
Given this information, we use a lattice representation for 
our DNA sequence, as illustrated in Figure \ref{ladder-fig}. 
%
%
Breathing -- the localised separation of complementary bases -- 
takes place on the microsecond timescale in normal DNA, which is 
beyond the range of all-atom molecular dynamics (MD) packages. 
The insertion of a defect, that is, replacing a thymine (T) with a 
difluorotoluene (F) base at the lattice site $n=0$, increases the 
frequency of breathing due to the weakening of the inter-chain 
potential. This makes breathing occur on the nanosecond timescale 
and hence it becomes accessible to MD simulation techniques.  
Biologically the reason for considering the inclusion of a defect 
is to study fidelity, and the effects of errors, in DNA replication.  
The defect F has been considered previously, for example, by 
Cubero {\em et al.}\ \cite{cubero}, who considered such a system 
without any externally imposed twist.  It is possible that proteins 
may locally alter the twist of a DNA helix in order to ease the 
process of localised melting. Hence, here, we impose a twist on the 
DNA structure in order to investigate its effect on base-pair breathing. 

In this paper, we analyse base-pair breathing at a defect in 
double-stranded DNA parameterised using the molecular dynamics 
(MD) simulation tool AMBER.    We consider a range of helicoidal 
twist angles from 30$^\circ$ to 40$^\circ$ per base-pair at rest, 
including the typical twist of 36$^\circ$. 
In \cite{Duduiala1} we derived a coarse-grained model based on 
stochastic differential equations with one variable for the displacement  
of each base from its equilibrium position.  This model includes 
explicit random forcing terms, which model the interactions of water 
molecules with the bases, and the whole model has been 
parameterised to the all-atom MD simulations derived using AMBER. 
Developed from the models of \cite{WattisLaughton,Wattis}, 
the incorporation of noise and damping terms enables us to 
investigate the temporal dynamics of breathing.   The model is fitted to 
AMBER data using a sophisticated maximum likelihood estimation 
procedure which is described in \cite{Duduiala1} and summarised in 
Section \ref{sde-sec} here.  The associated fluctuation-dissipation 
relation, the dependence of parameters on angle of twist and form of 
the inter-base potential are also discussed in \cite{Duduiala1}. 

The DNA sequence under study consists of 12 base-pairs and we have 
taken that the DNA sequence is surrounded by a water box.  
The analysis of the simulations of a DNA molecule obtained using 
AMBER and our SDE system presented in \cite{Duduiala1} revealed 
that the amplitude of fluctuations is slightly reduced in the SDE model, 
but the breathing length and frequency are similar. In addition, the 
derivation of the SDE model requires us to modify the 
fluctuation-dissipation relation in the reduced, or coarse-grained, 
mesoscopic models.  We also make a distinction between the 
potential of mean force (the free energy) and the various potential  
energies in our system.  This is supported by an analysis of the  
importance of the damping term in preserving the system energy  
(which, in combination with the forcing noise terms gives rise to  
the entropic component of the free energy) and the way in 
which the along-chain interactions influence the length of a breathing 
event. We have also shown in \cite{Duduiala1} that breathing events are 
due not only to inhomogeneities in the inter-strand interactions, but also 
to a significant change in along-chain interactions and the 
helical twist of the DNA, which potentially influences the interactions  
between the DNA molecule and the surrounding solvent. 

A more detailed analysis of our simulations presented below reveals 
an interesting result: rather than a well-defined breathing frequency 
that depends on the twist angle {\em via} the energy barrier between  
open and closed states of the central basepair, we find that at all  
angles of twist the DNA exhibits breathing across a wide range of 
frequencies and the amplitude-frequency relationship exhibits scale-free 
behaviour. Most previous results in the literature show that the energy 
is transferred between nonlinear localized modes with particular 
frequencies and between such modes and phonons, as discussed by 
Peyrard {\em et al.}\ \cite{Peyrard, PeyrardFarago, PeyrardLopezJames} 
or as shown by the results of Gaeta {\em et al.}\ \cite{Gaeta, Caldarelli, 
Cardoni}, for example. 

In the rest of this section we review some of the relevant theory and 
applications involving self-organised systems.  In Section \ref{model-sec} 
we outline the models we use and the simulation techniques.  
Section \ref{res-sec} contains the results of both approaches, 
showing the consistency of outcomes.  Section \ref{conc-sec} 
concludes the paper with a summary and discussion. 

\subsection{Self-organisation and criticality}

Many dynamical systems evolve to a steady state (or an equilibrium) 
solution or to a limit cycle. More complicated large-time behaviour 
includes spatio-temporal chaos which is often characterised by strange 
attractors in phase space \cite{Ruelle}. Through the study of cellular 
automata, Wolfram \cite{Wolfram} characterised large time behaviour into 
four classes. He empirically identified the following 
qualitative classes: spatially homogeneous systems (akin to 
steady-states), periodic structures (limit cycles), systems with chaotic 
aperiodic behaviour (chaos) and, most notably, a fourth, that of 
complicated localised and possibly propagating structures. 
We claim that this last category is the most appropriate classification 
for the behaviour which we observe later in this paper, and, more 
precisely, this behaviour corresponds to self-organised criticality (SOC).

In physics, a critical point specifies the conditions, such as temperature,
pressure or composition, at which a phase boundary is not valid anymore.
Here, by `phase' we understand a state of a system for which the physical
properties of a component are uniform. As one approaches the critical
point, the properties of the different phases approach each other.
In other words, a critical point refers to a system configuration to which 
the system evolves without ever approaching a fixed equilibrium state. 
Mathematically, one can define characteristic sizes of behaviour (lengths 
or timescales of events) which describe the system.  When these sizes 
become infinite the system is classed as critical, that is, fluctuations 
occur at all length scales.

Systems having the SOC property present a spatially or temporally 
scale-invariant behaviour without the need to tune any parameter to a 
specific value. This ubiquity of scale-free behaviour \cite{Zinn} in such 
models shows that complex behaviour can be stable. This contrasts 
with systems where one would generically expect some stable steady 
behaviour for a wide range of parameter values, and as a parameter 
changes, one might observe 
%
%
either a smooth change in the system's behaviour or a bifurcation to a 
qualitatively different steady-state or limit cycle, for example. 
In the special case 
when the parameter takes on the threshold value between two 
states, more complex phenomena may be observed.  This is a typical 
form of behaviour around a phase transition.  
In general, the total number of states is finite and the transitions can 
be characterised using a cellular automaton structure \cite{Chopard}.
       
When analysing a large system, we aim to reduce its 
complexity to a few degrees of freedom, for which the coupling 
can be defined in a general manner and hence we obtain some averaged, 
or coarse-grained, behaviour over ignored quantities and includes 
corresponding averaged interactions within the system and the 
surrounding environment.   For dynamical systems, such a dimensional 
reduction can be achieved by the  ``slaving 
principle'' \cite{Haken} which leads to the the study of low-dimensional 
attractors. This is often a straightforward method. For example, 
``fast modes'' at equilibrium can be slaved to a few slowly-evolving 
modes. However, sometimes a system responds on both fast and 
slow timescales, even at large times, and we require an alternative theory,
such as the idea of self-organised systems, whose behaviour cannot
be explained using the slaving principle or other reductions.
       
Some attracting critical points of dynamical systems can be
characterised using the concept of self-organized criticality (SOC),
which was first introduced by Bak {\em et al.}\ \cite{Bak87, Bak88}.
Using simple automata, they demonstrated power-law relationships and
$1/f$ noise (also known as ``flicker noise'') in spatially extended systems,
this behaviour illustrates critical phenomena, and underpins more
general scale-invariant behaviour and fractals.  They studied the 
dynamics of damped pendula and the slope of sandpiles, determining 
critical points of the systems. 
One aspect of self-organised criticality is the separation of timescales:
in the most familiar application of sandpiles, grains are continually added
on a faster timescale. The gradient of the pile slowly steepens
and there are avalanches (large-scale reorganisation of the pile) which
take place rapidly but are separated by large time intervals (relative to
the timescale at which grains are added to the pile).
Bak {\em et al} also noted that changing the values 
of system parameters did not affect the emergence of critical behaviour.
Avalanches in a one-dimensional sandpile are also analysed by Chapman
{\em et al.}\ (see, for example, \cite{Chapman2}) who showed that the
distribution of energy discharges due to internal reorganizations have a
power-law form and so demonstrate that the system is self-organized.

In general, for a noisy system the power spectrum has the form
$S(f)=cf^{-\beta}$, where $c$ is a constant. The noise present in 
the system can be classified in three important categories as follows:
\begin{itemize}
        \item white noise, for $\beta=0$;
        \item pink noise, for $\beta=1$;
        \item red noise (also known as Brownian noise), for $\beta=2$.
\end{itemize}
However, the term ``$1/f$ noise'' is widely used to refer to any noise
with a power spectral density $S(f)\propto f^{-\beta}$, with $0<\beta<2$.
For $1/f$ noise that occurs in nature, $\beta$ is usually close to 1.

Although there is no single well-defined class of systems having the SOC 
property, it is typically observed in complex systems with slowly-driven 
nonequilibrium behaviour, for which the causes of an event taking place 
in a system cannot be explained simply through some parameter values. 
Several studies of SOC show that scale-invariant phenomena can be 
determined at critical points, but not necessarily at any critical point. 
There are two important categories of such phenomena: fractals 
\cite{Peng} and power laws \cite{Newman}.  Whilst the first category 
involves geometric shapes, which can be split into parts that are 
reduced-size copies of the initial shape, the second deals with 
frequency-dependent quantities and, hence, is relevant in the analysis 
of some Hamiltonian systems.  We note that self-organised systems 
are always at criticality, but not all critical systems are self-organised.

The range of systems exhibiting critical properties varies from earthquakes
\cite{Olami, BakEarth} and forest-fires \cite{BakFF, Drossel} to biological
systems, such as proteins \cite{Phillips, Phillips2}, the brain \cite{Werner}
and even DNA. Selvam \cite{Selvam2}, for example, studies the distribution
of bases in a human DNA sequence and shows that the C-G base-pair
frequency distribution exhibits a universal inverse power-law form.
Also, Harris {\em et al.}\ \cite{Harris} analyse the configurational entropy
of a DNA molecule based on the entropy estimation for a Gaussian
configuration given by Schlitter \cite{schlitter}, which helps investigate
whether a steady state has been reached during a simulation. They
show that the estimate of the entropy $S_{n}$ depends on the number
of data points $n$ and this relation is a power law (with exponent 
between zero and minus one). 

\section{Modelling DNA}
\label{model-sec}

In our system, we also have a separation of timescales:
there are rapidly oscillating forcing terms applied to the DNA chain
illustrated in Figure \ref{ladder-fig} (these model the interactions with 
water molecules and are akin to grains being added
to the sandpile) and the occasional larger-scale restructuring of
the chain as the base-pairs open or close at the start and end of
breathing events (which are akin to avalanches in sandpiles).
We now discuss in detail the modelling approaches we adopt.

\subsection{All-atom MD modelling using AMBER}
\label{amber-sec}

As already mentioned, we obtain data from an all atom simulation of 
DNA using the package AMBER \cite{Amber}. The DNA sequence 
analysed contains 12 base-pairs as follows:
\begin{verse}
	\begin{tabbing}
		C\ \ \=\ \ T\ \ \=\ \ T\ \ \=\ \ T\ \ \=\ \ T\ \ \=\ \ G\ \ \=\ \ F\ \ \=\ \ A\ \ \=\ \ T\ \ \=\ \ C\ \ \=\ \ T\ \ \=\ \ T\  \\
	  G\ \ \>\ \ A\ \ \>\ \ A\ \ \>\ \ A\ \ \>\ \ A\ \ \>\ \ C\ \ \>\ \ A\ \ \>\ \ T\ \ \>\ \ A\ \ \>\ \ G\ \ \>\ \ A\ \ \>\ \ A
	\end{tabbing}
\end{verse}
This sequence is analysed at a constant temperature of $T = 293K$, 
in the presence of a surrounding water box. The solvent has to be taken 
into account because it influences the displacement of atoms and through 
other bonds, affects the hydrogen bonds linking the bases on the DNA 
strands \cite{PeyrardLopez}. Even if the breathing events occur on the 
nanosecond time-scale and the DNA sequence contains only 
12 base-pairs, which together with the sugars and phosphate groups 
represent 763 atoms, the number of degrees of freedom in our system 
is actually very large (16682) due to the water box. This means most 
of the time is spent computing information about the solvent, even 
though this information is not used for our analysis, since we focus 
only on the DNA bases and their dynamics. 
	
Note that AMBER considers that the normal twist by default is about 
32.5$^\circ$. In order to avoid this inconvenience, we have constructed 
the DNA sequence by considering the degree of twist at rest. The degree 
of twist is preserved by imposing a harmonic restraint 
on the atoms at the end bases. More precisely, we have considered 
a constant energy (of 1 kcal mol$^{-1}$ \AA{}$^{-2}$) and hence a 
constant force acting on the end bases, in order to keep the DNA 
atoms close to their initial positions. Applying this restraint 
to the end bases allows the A-F pair to breathe and so explore 
a larger volume of phase space than the other base-pairs.
	
The force field used during the AMBER simulations is also important. 
For a DNA molecule, the predefined FF99SB all-atom force field was 
used. Moreover, AMBER provides several water models for residues 
with name WAT -- the default is TIP3P, which was used in our simulations. 
After creating the topology and coordinates files using the AMBER 
packages \textit{LEaP} and \textit{nucgen}, we have used \textit{SANDER} 
for energy minimization. This process involves a structural relaxation, 
which is necessary because the coordinates file contains some initial 
values that do not guarantee a minimum energy configuration, this 
reduces the possibility of having conflicts or overlapping atoms. In 
addition to energy minimization, we have also performed a few 
equilibration and MD simulations in which temperature changes.
	
Finally, our system is simulated using \textit{SANDER}. Next, 
\textit{ptraj} is used to measure the distance between the A-F 
base-pair as well as the separations of the other base-pairs and 
the corresponding velocities. This data is output every 1 ps or 
2 fs depending on which simulation study is being performed. 

\subsection{Stochastic mesoscopic model}
\label{sde-sec}

\begin{figure}[!ht]
\begin{center}
\includegraphics[scale=0.65]{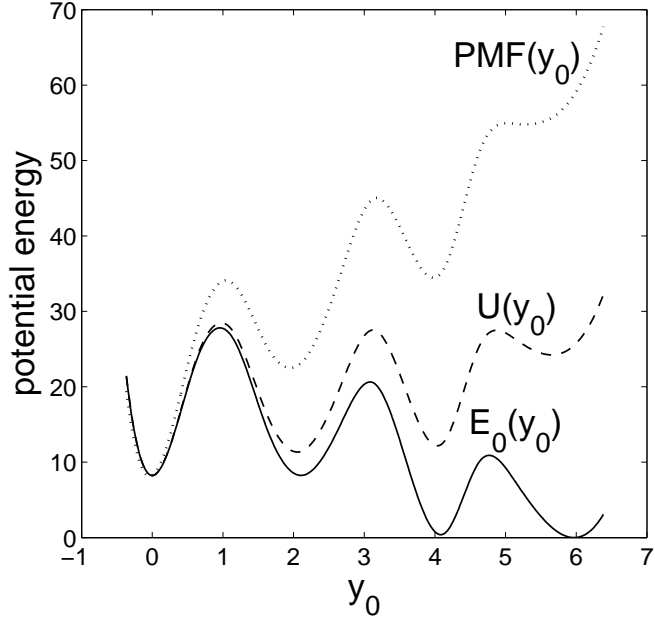}
\end{center}
\caption{The interchain interaction potential $E_0(y_0)$,
the total potential energy $U(y_0)=E_0(y_0)+\frac{1}{2}\hat{k}y_0^2$,
and the potential of mean force $PMF(y_0) = U(y_0) + \sqrt{k_B T} 
\eta_0 y_0$ used in (\ref{SYSWNM_3}), for a 36$^{\circ}$ twisted 
DNA sequence.}
\label{E0_FIG}
\end{figure}

In DNA, the rapid external fluctuation events correspond to random 
collisions of the bases with the water molecules surrounding the 
biopolymer.  These events correspond to perturbations to the variables 
$y_n(t)$, fluctuations in these displacements may propagate along the 
chain, perturbing neighbouring bases, eventually causing the central base 
$y_0(t)$ to cross a barrier in the potential $E_0(y_0)$, see
Figure \ref{E0_FIG}. 

A second source of data is the reduced model introduced in 
\cite{Duduiala1} which is based on a system of stochastic ordinary 
differential equations (SDEs). We start from the deterministic system
\begin{eqnarray}
H & = & \sum_n \left\{ \half \dot u_n^2 + \half \dot v_n^2
+ \half \gamma (u_n-v_n)^2
+ \half k (u_{n+1}-u_n)^2 + \half k (v_{n+1}-v_n)^2 \right\} \nn\\ &&
+ \half (\hat k -k )  \left[ (u_1-u_0)^2 + (u_0-u_{-1})^2
+ (v_1-v_0)^2 + (v_0-v_{-1})^2 \right] \nn \\ &&
+ E_0(y_0) - \half \gamma (u_0-v_0)^2 , 
\end{eqnarray}
which is illustrated in Figure \ref{ladder-fig}.  Here, $u_n(t)$ represent the 
deviations from equilibrium of the bases on one chain of the DNA, and 
$v_n(t)$ the displacements of the bases on the second chain, with the 
index $n$ determining the position down each chain ($-6 \leq n \leq 5$). 
Since we are interested in investigating in more detail a system similar 
to that of Guckian {\em et al.}\ \cite{Guckian} we place a defect at the 
centre of the chain, that is, the $u_0$ and $v_0$ bases correspond to 
the A-F base-pair. The parameter $k$ describes the stiffness of the 
backbone down each side of the double-helix structure, whist the 
parameter $\gamma$ indicates the strength of interaction between a 
base on one strand and its complement.  These forces are assumed to be
uniform along the DNA double helix, except at the defect where
they are replaced by the parameter $\hat k$ and the nonlinear
force $E_0'(y_0)$ respectively. To these we add noise and damping
terms with coefficients $\epsilon_*$ and $\eta_*$ respectively.
The dependence of all these parameters on the twist angle
$\theta$ has been determined in an earlier paper \cite{Duduiala1}. 
We make the transformation $y_n=u_n-v_n$ so as to obtain
equations of motion for the distances between base-pairs
\begin{eqnarray}
\!\frac{\dd^{2}\!y_{n}}{\dd t^{2}} \!&\!=\!&\!
        k(y_{n+1}\!-\!2 y_{n}\!+\!y_{n-1}) - \gamma y_{n}
        - \eta\frac{dy_{n}}{dt} + \epsilon\xi_{n}, \quad ( |n|>1), 
        \label{SYSWNM_1} \\
\!\;\; \frac{\dd^{2}\!y_{-1}}{\dd t^{2}} \!&\!=\!&\!
        \hat{k}(y_{0}\!-\!y_{-1})-k(y_{-1}\!-\!y_{-2})- \gamma y_{-1}
        - \eta\frac{\dd y_{-1}}{\dd t} + \epsilon\xi_{-1},
        \label{SYSWNM_2} \\
\!\frac{\dd^{2}\!y_{0}}{\dd t^{2}} \!&\!=\!&\!
        \hat{k}(y_{1}\!-\!2 y_{0}\!+\!y_{-1})- \frac{\dd E_{0}}{\dd y}(y_{0}\!)
        - \eta_{0}\frac{\dd y_{0}}{\dd t} + \epsilon_{0}\xi_{0},
        \label{SYSWNM_3} \\
\!\frac{\dd^{2}\!y_{1}}{\dd t^{2}} \!&\!=\!&\!
        k(y_{2}\!-\!y_{1})-\hat{k}(y_{1}\!-\!y_{0})- \gamma y_{1}
        - \eta\frac{\dd y_{1}}{\dd t} + \epsilon\xi_{1}. \label{SYSWNM_4}
\end{eqnarray}
The functions $\xi_n(t)$ are white noise forcing terms. The quantities
$\gamma$, $k$, $\hat{k}$, $\epsilon$, $\epsilon_0$, $\eta$, $\eta_0$ are
all fitted using a maximum likelihood estimation procedure, as is the
interaction potential $E_0(y_0)$, a full description of this is given in
\cite{Duduiala1,Duduiala4}.  A typical example of the multiwelled
interaction potential, $E_0(y_0)$ is given in Figure \ref{E0_FIG}. Note
that the total potential energy of the central defective base-pair is
$U(y_0) = E_0(y_0) + \frac{1}{2} \hat{k}y_0^2$, and the potential of
mean force (free energy) is given by $PMF(y_0) = U(y_0) +
\sqrt{k_B T}\eta_0 y_0$, can easily be obtained from simulations 
by plotting a histogram of binned displacement data.

\section{Results}
\label{res-sec}

We measure the distance between the two bases (A, F) at the defect 
over many nanoseconds, that is, we sample $y_0(t)$ at specific 
intervals.  Using data from both AMBER and the SDE system we 
analyse the (discrete) Fourier transforms of $y_0(t)$ for a range of twist 
angles from 30$^\circ$ to 40$^\circ$ per base-pair.  

The discrete Fourier transform is taken of the data, in the figures displayed 
later, we use the notation $DFT(\omega)$ for $\widehat y_0(\omega)$, 
and the hypothesis we are testing is that over a wide range of $\omega$, 
\begin{equation} 
\log \widehat y_0(\omega) = -\beta \log \omega + \log C .  
\label{loglin} 
\end{equation} 
The highest frequency attainable is $\omega_{{\rm max}} = 2\pi/2\Delta t$, 
whilst the lowest frequency is given by $2\pi / T$ where $T$ is the length 
of the simulation and $\Delta t$ is the sampling interval.   Our initial results 
are simulations of length approximately $T=10$ ns sampled every $\Delta t 
=1$ ps (giving $\sim 10^4$ data points).  We then perform simulations 
of $T=2$ ns sampled on a much finer scale of $\Delta t=2$ fs (giving 
$\sim10^6$ data points).   These are performed using both AMBER and 
our SDE system.  The SDE system is then subjected to a longer-time 
simulation of $T=100$ ns, sampled every $\Delta t=1$ ps ($\sim10^5$ data 
points). This enables a wide range of frequencies to be sampled: for the 
initial simulations $0.00063 < \omega < 3.14$ ps$^{-1}$ 
($-7.4 < \log \omega < 1.14$), for the more frequently sampled simulations 
$0.0032 < \omega < 1600$ ($-5.7 < \log \omega < 7.4$), and for the long 
simulations $2\pi\times10^{-5}<\omega<3.14$ ($-9.7<\log\omega<1.14$). 

Other frequencies which might be of relevance in interpreting the results 
are the upper and lower limits of the phonon band (which we refer to as 
$\omega_{{\rm opt}}$ and $\omega_{{\rm ac}}$ respectively), and the 
frequency of the defect mode. Assuming that the defect mode can be 
approximated by $\ddot{y}_0 = -E_0''(0) y_0$ we find $E_0''(0) = 14$ 
and $\omega_{{\rm def}}=3.7$, hence $\log\omega_{{\rm def}}=1.3$. 
Since $120 < \gamma < 165$ and $160 < \gamma + 4 k < 210$, 
we have $\omega_{{\rm ac}} = 11$ and $\omega_{{\rm opt}} = 14.5$, 
implying $\log \omega_{{\rm ac}} = 2.4$ and 
$\log \omega_{{\rm opt}} =2.7$.   These all lie well above 
the range of frequencies that we shall be interested in below. 

\subsection{Initial results}
\label{init-sec}

\begin{figure}[!ht]
\begin{center}
\includegraphics[width=90mm,height=120mm]{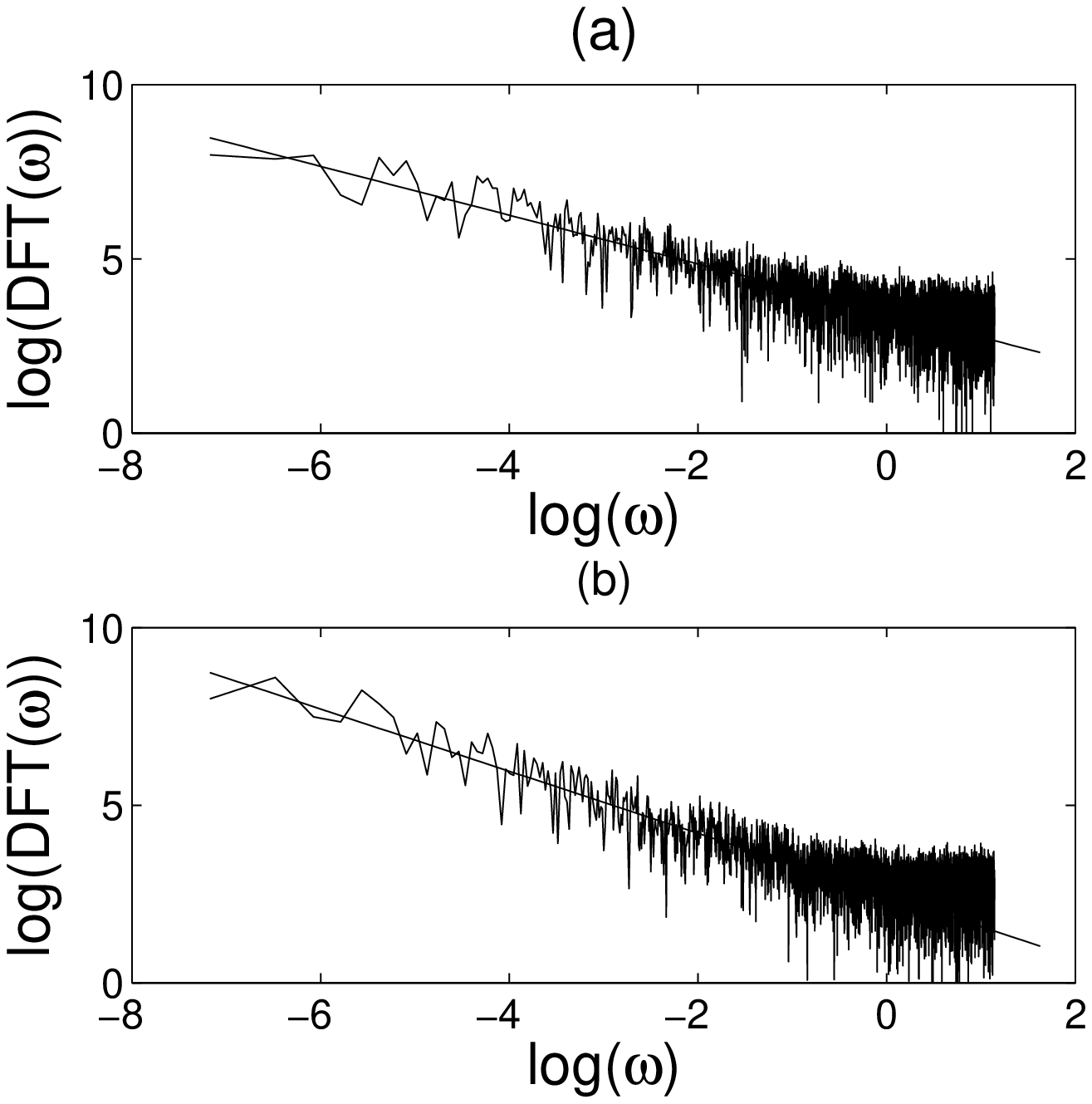}
\end{center}
\caption{Discrete Fourier transforms (power spectra) of 
$y_0(t)$ plotted against $\omega$ on a log-log scale. 
Data sampled every 1 ps for 10 ns from a simulation 
of a 38$^{\circ}$ twisted DNA sequence from (a) AMBER,
and (b) the SDE system (\ref{SYSWNM_1})--(\ref{SYSWNM_4}). }
\label{FT_FIG}
\end{figure}

\begin{figure}[!ht]
\begin{center}
\includegraphics[height=120mm,width=90mm]{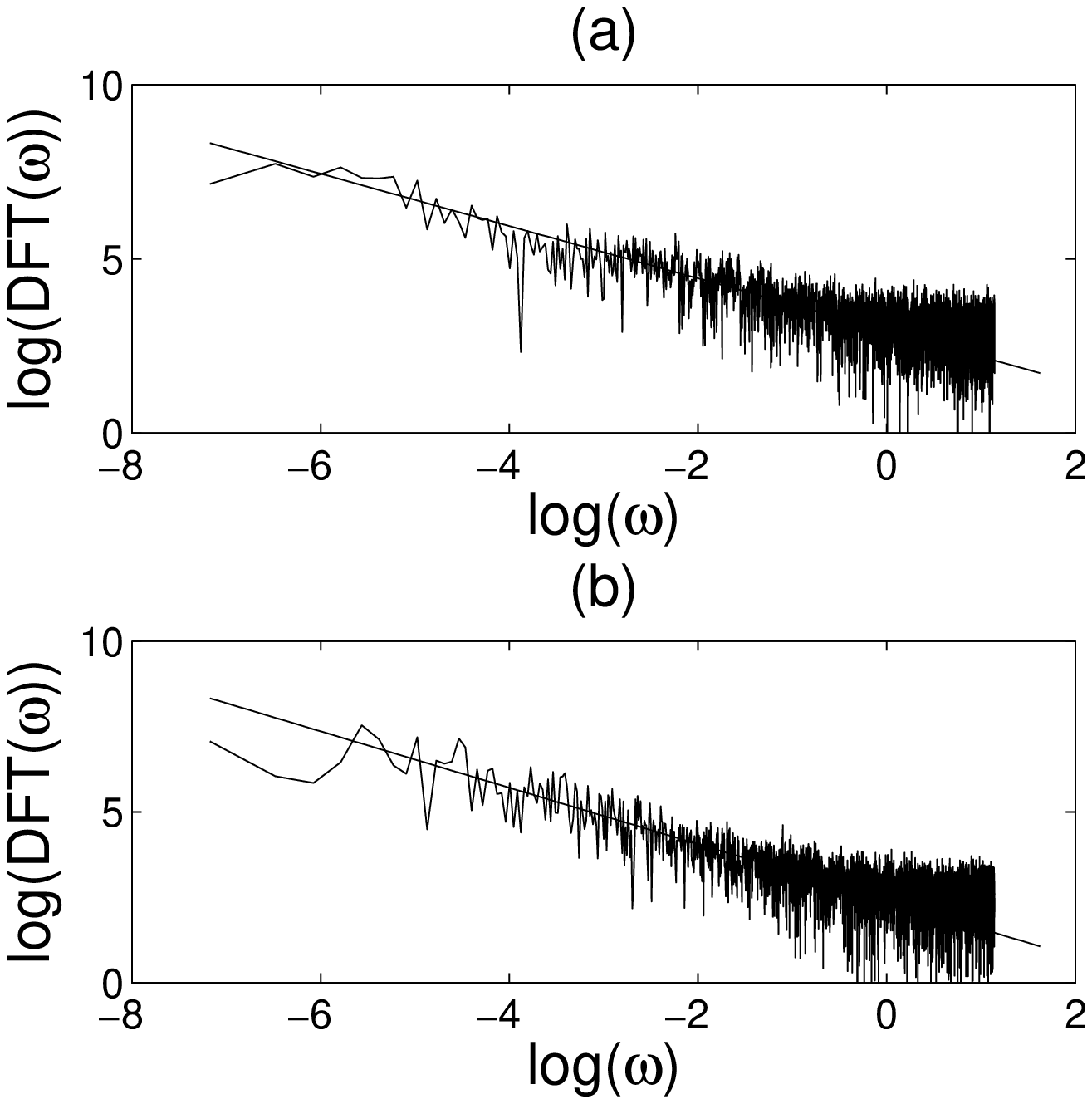}
\end{center}
\caption{Discrete Fourier transforms (power spectra) of $y_0(t)$ 
plotted against $\omega$ on a log-log scale. Data sampled every 
1 ps for 10 ns from a simulation of a 34$^{\circ}$ twisted 
DNA sequence from (a) AMBER, and (b) the SDE system 
(\ref{SYSWNM_1})--(\ref{SYSWNM_4}). }
\label{FT_FIG_34}
\end{figure}

In Figure \ref{FT_FIG}, we present the log-log plot of the discrete 
Fourier transform (DFT) against the frequency $\omega$, for $2^{13}$ 
data points, that is, $y_0(t)$ sampled every 1 ps for 8 ns, for a 
38$^{\circ}$ overtwisted DNA sequence. 
This Figure shows a straight line fit over several orders of magnitude 
of $\omega$ ($-7<\log_e\omega<-1$), with gradients of 0.7--0.9 which 
are close to $-1$.  Similar results are obtained for a 34$^{\circ}$ twisted 
DNA sequence, for example, as can be seen in Figure \ref{FT_FIG_34}.
These results suggests that there are breathing events of, and separated 
by, arbitrarily large times.   Thus, if a DNA strand was successively 
observed for increasingly long intervals of time, there would always be 
breathing events of duration comparable to the total observation time. 
The gradients of the $\log_e$DFT($\omega$) against 
$\log_e\omega$ lines for the full range of twist angles tested 
are summarised in Table~\ref{BETAC_VAL} and suggest the 
presence of generalised $1/f$ noise in our data (that is, 
$1/f^\beta$ with $0<\beta<2$). 

\begin{table}[!ht]
\begin{center}
\begin{tabular}{|| c | c | c ||} \hline\hline
Angle & $\beta_{AMBER}$ & $\;\;\beta_{SDE}\;\;$ \\ \hline
30$^\circ$               & 0.725  &  0.750      \\ \hline
32$^\circ$               & 0.700  &  0.725      \\ \hline
33$^\circ$               & 0.725        &  0.775        \\ \hline
34$^\circ$               & 0.750        &  0.825        \\ \hline
35$^\circ$               & 0.750        &  0.775        \\ \hline
36$^\circ$               & 0.775        &  0.825        \\ \hline
38$^\circ$               & 0.700        &  0.875        \\ \hline
40$^\circ$               & 0.700        &  0.700        \\ \hline\hline
\end{tabular}
\end{center}
\caption{The gradient $\beta$ of the log-log representation of 
$DFT(y_{0})$ against $\omega$, from $\sim10$ ns of data, sampled 
every 1 ps obtained from the AMBER and SDE models.}
\label{BETAC_VAL}
\end{table}

Initially our aim was to identify the dominant frequencies of breathers at 
the defect through the interchain distances. 
The asymptotic results of \cite{Wattis} initially appear to suggest that 
breathers are time-periodic modes with well-defined frequencies; 
however, that theory actually predicts a one-parameter family of 
``in-phase'' breather modes with frequencies occupying the full range of 
values from the bottom of the phonon band down to arbitrarily small 
frequencies (as well as a one-parameter family of ``out of phase'' breathers 
with frequencies above the top of the phonon band).   Since the part of 
the frequency range that we are interested in here is the small-$\omega$ 
limit, it is the former, in-phase, family that concerns us here. 
We assume that a combination of the stochastic forcing noise and 
nonlinear interactions of phonons with the breathers which changes 
their frequency over time and may even sporadically create and destroy 
the breather modes. 

\subsection{More refined results}
\label{refined-sec}

\begin{figure}[!ht]
\begin{center}
\includegraphics[height=120mm,width=90mm]{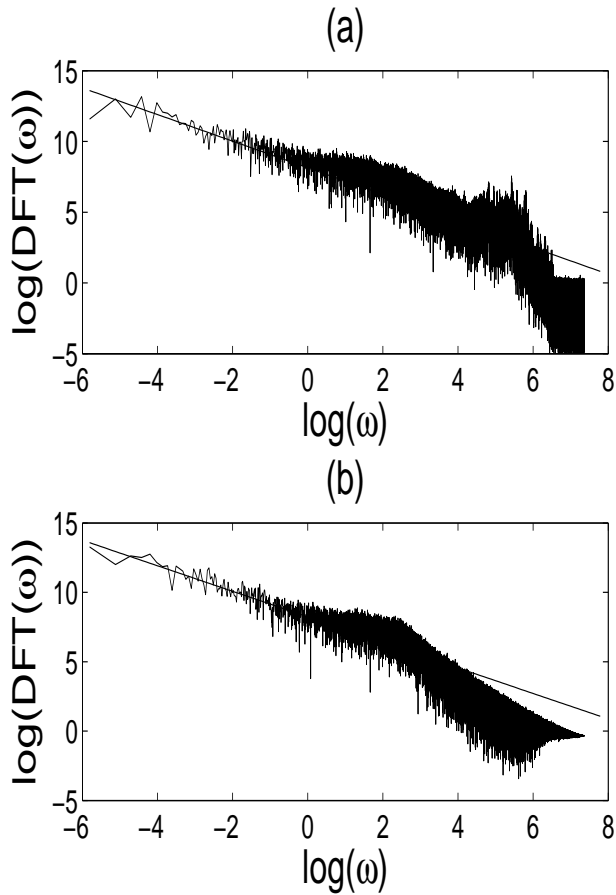}
\end{center}
\caption{Log-log plots of the discrete Fourier transforms 
(power spectrum) from more refined samplings of $y_0(t)$, 
specifically, data was sampled every 2 fs for 2.1 ns. 
Results for a 34$^{\circ}$ 
undertwisted DNA sequence from (a) AMBER and (b) the SDE 
system (\ref{SYSWNM_1})--(\ref{SYSWNM_4}). } 
\label{FT_FIG34}
\end{figure}

By decreasing the sampling interval ($\Delta t$), we expect to obtain more 
accurate results; the cost being the increase in data storage requirements.  
Analysing data sampled every 2 fs for 2.1 ns (more precisely, $2^{20}$ or 
$10^6$ data points), we obtain qualitatively similar results, as can be seen 
in the results presented in Figure \ref{FT_FIG34}, which shows the Fourier 
power spectra for a DNA helix twisted to 34$^\circ$.  

As with the results shown in Section \ref{init-sec}, the full range of 
twist angles from 30$^\circ$ to 40$^\circ$ have been simulated and 
analysed the results are summarised in Table~\ref{BETA_VAL}. 
These results, with the reduced sampling interval, suggest that the 
average value of $\beta$ for such AMBER simulations increases to 
$0.93$, which is much closer to unity than the values obtained when 
the sampling interval is $\Delta t=1$ps (Table~\ref{BETAC_VAL}). 
Analysing a similar data set from the SDE model with more frequent 
sampling shows a similar increase in $\beta$, from an average of $0.79$
when $\Delta t=1$ps (Table~\ref{BETAC_VAL}) to $\Delta t = 0.91$
(Table~\ref{BETA_VAL}). These results again confirm that our reduced 
mesoscopic model accurately reproduces the self-organised DNA 
behaviour observed in the all-atom AMBER simulation.   This observed 
increase in $\beta$ suggests that we might reasonably expect 
$\beta\rightarrow 1$ as $\Delta t\rightarrow0$ in both the AMBER 
and the SDE simulations.

\begin{table}[!ht]
\begin{center}
\begin{tabular}{|| c | c | c ||} \hline\hline
Angle            & $\beta_{AMBER}$ & $\;\;\beta_{SDE}\;\;$\\ \hline
30$^\circ$               & 0.920  & 0.900\\ \hline
32$^\circ$               & 0.920  & 0.895\\ \hline
33$^\circ$               & 0.900  & 0.910       \\ \hline
34$^\circ$               & 0.940  & 0.920       \\ \hline
35$^\circ$               & 0.930  & 0.900       \\ \hline
36$^\circ$               & 0.930  &     0.910 \\ \hline
38$^\circ$               & 0.940  & 0.940       \\ \hline
40$^\circ$               & 0.950  & 0.905       \\ \hline\hline
\end{tabular} \end{center}
\caption{The $\beta$ exponents derived from more refined 
sampling of $y_0(t)$, specifically every 2 fs for 2.1 ns.
The $\beta$-exponent is the gradient of the log-log 
representation of $DFT(y_{0})$ against $\omega$, from 
AMBER and SDE simulations (compare with Table \ref{BETAC_VAL}). }
\label{BETA_VAL}
\end{table}

At higher values of $\omega$, both AMBER and SDE simulations show 
changes in behaviour.   Firstly, in both cases, the line broadens as 
more points are plotted at higher values of $\log \omega$ 
and these display a greater variation;  see the ranges $-1<\log\omega<2$ 
in Figures \ref{FT_FIG} and \ref{FT_FIG_34} and $-1<\log\omega<6$ in 
Figure \ref{FT_FIG34}.  Note that the same ranges apply to both 
AMBER simulations and SDE simulations. 
Secondly, at larger $\omega$, there is a more significant reduction 
in the discrete Fourier transform, for the AMBER simulation this occurs 
from $\omega=2$ upwards, following by a more abrupt decrease around 
$\omega=6.5$ in Figure \ref{FT_FIG34}(a); whereas, in the SDE simulation, 
Figure \ref{FT_FIG34}(b), the spectrum has a simpler form with a more 
rapid linear decrease with a larger gradient beyond $\omega=2.5$.  

\subsection{Long-time results}
\label{long-sec}

\begin{figure}[!ht]
\begin{center}
\includegraphics[scale=0.65]{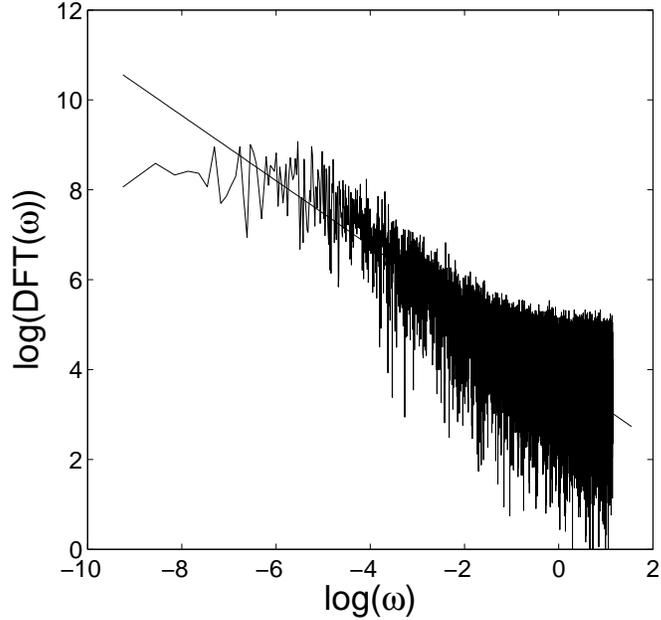}
\end{center}
\caption{Log-log plot of the discrete Fourier transform 
(power spectrum) of $y_0(t)$ from a long-time (100 ns) SDE
simulation for a 30$^\circ$ twisted DNA sequence, 
$y_0(t)$ data sampled every 1 ps.}
\label{LOGFFT_FCT30_LONG}
\end{figure}

\begin{figure}[!ht]
\begin{center}
\includegraphics[scale=0.65]{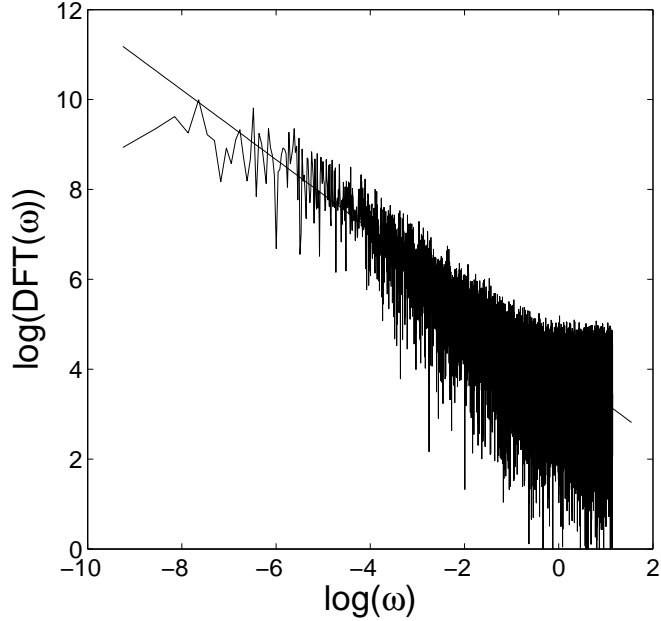}
\end{center}
\caption{Log-log plot of the discrete Fourier transform 
(power spectrum) of $y_0(t)$ from a long-time (100 ns) SDE
simulation for a 38$^\circ$ twisted DNA sequence, 
$y_0(t)$ data sampled every 1 ps.}
\label{LOGFFT_FCT38_LONG}
\end{figure}

We have analysed a longer simulation of 100 ns, sampling $y_0(t)$ 
every $\Delta t=$ 1 ps using just the SDE system; such a long simulation 
is beyond the scope of AMBER on currently available computing facilities.  
This length of simulation allows lower breathing frequencies to be 
sampled, as shown in Figures \ref{LOGFFT_FCT30_LONG} and 
\ref{LOGFFT_FCT38_LONG}.  Here we observe the same self-organised 
behaviour as in earlier graphs, although with some deviation from the 
straight line at particularly small frequencies, namely those in the range 
$-9 < \log \omega < -7$. 

For example, from the $2^{16} (\sim 10^5)$ data points in the simulation 
of a 30$^{\circ}$ undertwisted DNA sequence illustrated in Figure 
\ref{LOGFFT_FCT30_LONG}, we find $\beta=0.725$. This value is similar 
to that found in the shorter simulation of 10 ns sampled every 1 ps, 
where we found $\beta=0.750$; the value of $0.725$ is identical to the 
value obtained from the shorter AMBER simulation; both these values 
are reported in Table \ref{BETAC_VAL}. 

For the 38$^{\circ}$ overtwisted DNA molecule the long-time SDE 
simulation gives $\beta=0.775$, which lies between the value of 
$\beta=0.875$ from the shorter SDE simulation and $\beta=0.700$ 
from the shorter AMBER simulation.   Thus for both twist angles, the 
long-time SDE simulation gives exponents closer to the AMBER 
results than the shorter SDE simulations. 

\subsection{Bases away from the defect}

\begin{figure}[!ht]
\begin{center}
\includegraphics[width=90mm,height=120mm]{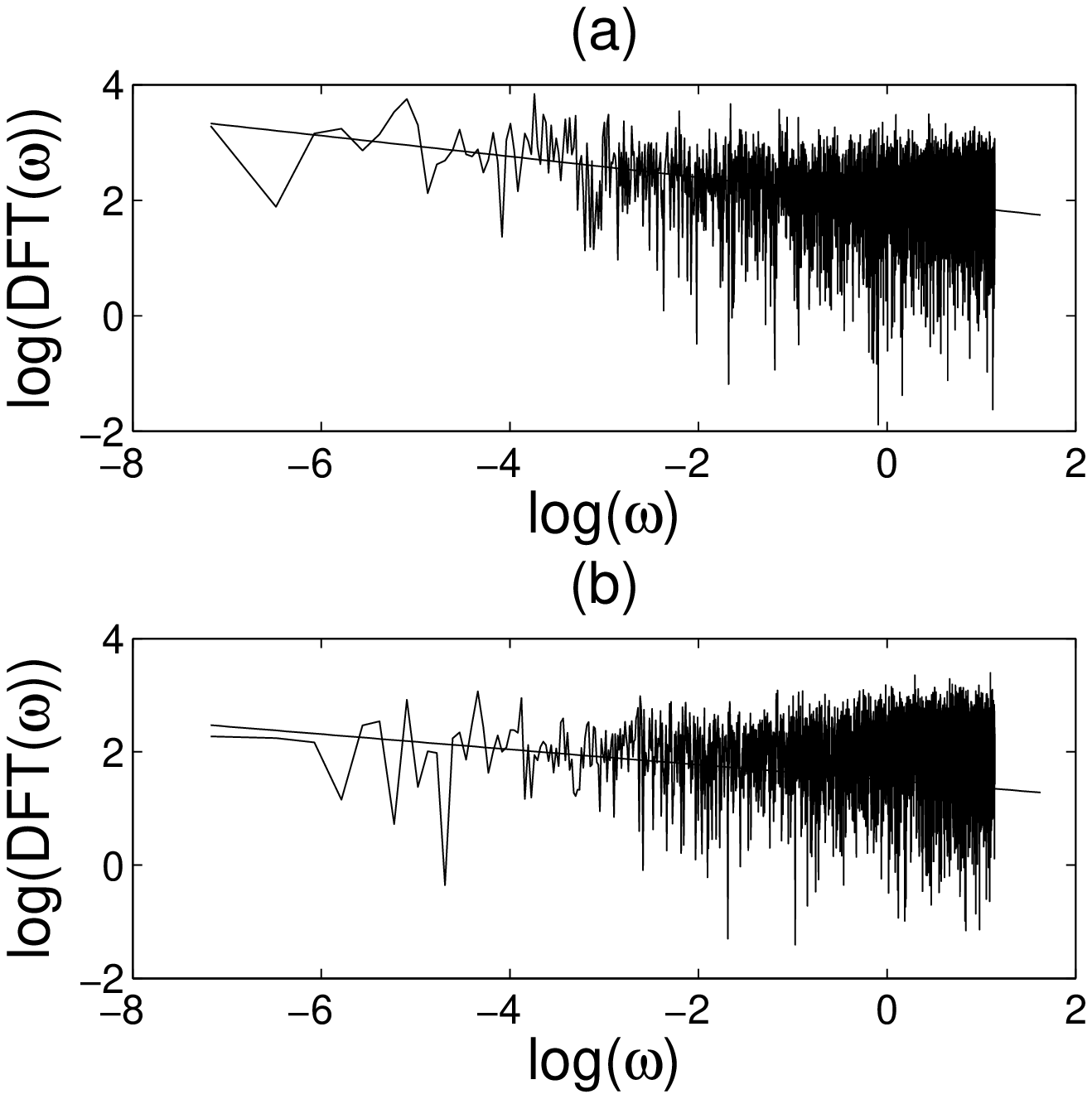}
\end{center}
\caption{Discrete Fourier transforms (power spectra) of $y_2(t)$ 
plotted against $\omega$ on a log-log scale. Data sampled every 
1 ps for 10 ns from a simulation of a 38$^{\circ}$ twisted DNA 
sequence from (a) AMBER, and (b) the SDE system 
(\ref{SYSWNM_1})--(\ref{SYSWNM_4}).}
\label{LOGFFT_Y2}
\end{figure}

Finally we have analysed the motion of the base-pairs adjacent and 
further away from the defect in the AMBER and SDE systems, in 
both cases using the example of a 38$^\circ$ overtwisted DNA helix. 
For example, Figure \ref{LOGFFT_Y2} illustrates the power spectrum 
of the second-neighbour base-pair $y_2{t}$, sampled every 1 ps over 
a simulation of length 10ns.  
Although the decay with increasing frequency ($\omega$) is not 
as clear as in Figures \ref{FT_FIG}, \ref{FT_FIG_34}, \ref{FT_FIG34}, 
\ref{LOGFFT_FCT30_LONG}, or \ref{LOGFFT_FCT38_LONG}, it is 
still possible to fit a straight line through the points.  For the AMBER 
and SDE simulations respectively, the gradients of these lines are 
0.225 and 0.210 respectively.  

This procedure has been repeated for the first neighbour base-pair, 
$y_1(t)$, and more distant base-pairs, $y_3(t)$, $y_4(t)$, and the 
corresponding gradients of the best fit lines of the log-log plots have 
been calculated.  Assuming that the FTs have power-law forms,  these 
gradients, which correspond to the exponents, $\beta$, are given in 
Table~\ref{BETA_VAL_NBR_AMBER}.   From the data in this Table we 
observe that $\beta$ decreases as one moves away from the defect 
site.  This behaviour is due to the reduced influence of the breathing pair 
on the neighbouring base-pairs. We observe a drop from $\beta\approx1$ 
at the defect ($n=0$) to just under one quarter at the nearest 
neighbour ($n=1$) in both SDE and AMBER.

\begin{table}[!ht]
\begin{center}
\begin{tabular}{|| c | c | c ||} \hline\hline
Base-pair  & $\beta_{AMBER}$ & $\;\;\beta_{SDE}\;\;$ \\ \hline
$y_{1}(t)$               & 0.225  & 0.210 \\ \hline
$y_{2}(t)$               & 0.180  & 0.135 \\ \hline
$y_{3}(t)$               & 0.180  & 0.130 \\ \hline
$y_{4}(t)$               & 0.180  & 0.130 \\ \hline\hline
\end{tabular}
\end{center}
\caption{The gradient $\beta$ of the log-DFT function,
from AMBER and SDE data sampled every 1 ps for 10 ns.}
\label{BETA_VAL_NBR_AMBER}
\end{table}

\begin{table}[!ht]
\begin{center}
\begin{tabular}{|| c | c | c ||} \hline\hline
Base-pair     & $\beta_{AMBER}$ & $\;\;\beta_{SDE}\;\;$ \\ \hline
$y_{1}(t)$               & 0.450  &  0.445\\ \hline
$y_{2}(t)$               & 0.425  &  0.205\\ \hline
$y_{3}(t)$               & 0.410  &  0.180\\ \hline
$y_{4}(t)$               & 0.390  &  0.155\\ \hline\hline
\end{tabular} 
\end{center}
\caption{The gradient $\beta$ of the log-DFT function,
from AMBER and SDE data sampled every 2 fs for 2 ns. }
\label{BETA_VAL_NBR}
\end{table}

The analysis of $y_0(t)$ showed that the exponent $\beta$ increased 
from $0.7-0.8$ to $0.90-0.95$ when the sampling frequency was 
decreased from 1ps to 2 fs, suggesting convergence to $\beta=1$ 
in the limit of small sampling frequency.  Hence, we attempt to find 
more accurate values for the $\beta$-exponent for the neighbouring 
bases by decreasing the sampling timestep from 1 ps to 2 fs (as we 
reduce the simulation length from 10ns to 2ns).   We obtain the 
gradients given in Table \ref{BETA_VAL_NBR}. Summarising, we find 
values just under one half at the nearest neighbour ($n=1$) in both SDE 
and AMBER.  For $y_2$, in AMBER, there is then a further slow decay 
of ${\cal O}(0.01)$ per base-pair; whereas in SDE there is a more 
significant drop to 0.2 for $y_2$ and then a slow decrease of 
${\cal O}(0.01)$ per base-pair.
We observe once again that these later results with a decreased sampling 
timestep of $\Delta t= 2$ fs increases the measured values of $\beta$. 
Hence, we recommend that the even with spatially coarse-grained models 
of DNA, kinetic characteristics should be analysed using as small a 
timestep as possible, even as small as 2 fs (as used in AMBER) in order 
to obtain correct results and conclusions.

This extra decrease in the SDE system may be due to the reduced 
number of degrees of freedom (only one per base-pair in the SDE),
whereas in AMBER there are ${\cal O}(10^2)$ degrees of freedom 
per base-pair (the bases having 15 atoms moving in 3D space, 
in addition to the phosphate backbone).  Whilst all base-pairs 
receive energy in the form of white noise forcing (in the SDE system) 
and in the form of random collisions with water molecules (AMBER), 
this energy is used and dissipated differently in the defective base 
from its neighbours.  At the defect, there is a change in temporal 
behaviour, since white input noise ($\xi_0$) is converted to pink output 
($y_0$) giving $\beta\approx 1$, whereas in the neighbouring 
bases, the output noise ($y_n$) remains significantly closer to white, 
that is $\beta$ is significantly smaller.

\section{Conclusions}
\label{conc-sec}

In summary, we have simulated a sequence of 12 base-pairs of DNA
using two different models for a variety of times up to 100ns. 
We have analysed the displacement between the defective base-pair 
near the centre of the chain.
Fourier transforms of the distance trajectory taken from both the AMBER
and the mesoscopic stochastic differential equation simulations exhibit 
scale-free, or critical, behaviour for all twist angles in the range
30$^\circ$-40$^\circ$ per base-pair.   From (\ref{loglin}), we have 
$\widehat y_0(\omega) = C \omega^{-\beta}$ across a considerable 
range of frequencies, $\omega$.  Furthermore, we find that $\beta=1$ 
for all twist angles.  
Although we have imposed white noise forcing in the system
(\ref{SYSWNM_1})--(\ref{SYSWNM_4}), the noise observed in 
the output $y_0(t)$ is pink ($\beta\approx1$, Figure \ref{FT_FIG}).
This shows that our SDE system preserves the SOC properties 
of DNA observed in the fully deterministic all-atom AMBER simulations. 

It is suspected that proteins which interact with DNA overtwisting or 
undertwisting the structure in order to ease the release of bases out 
from the structure. The fact that $\beta=1$ for all twist angles appears 
to suggest that such a strategy will not change the base-pair breathing. 
However, the constant $C$ in the formula $\widehat y_0(\omega) = 
C \omega^{-\beta}$ will depend on twist angle, and mean that the 
fraction of time spent in the breathing state is less for the more stable 
angles $35^\circ-36^\circ$ and more for the overtwisted or 
undertwisted DNA structures ({\em i.e}\ those in the ranges 
$38^\circ-40^\circ$ and $30^\circ-34^\circ$ respectively).  

What is surprising about our results is that the critical behaviour is not 
specific to any one twist angle but occurs at all angles. One might expect 
that, at normal twist angles of 35$^\circ$--36$^\circ$, stable behaviour 
would be observed, with breathing events having some short 
characteristic timescale; at smaller and larger twist angles, a critical point 
would be found, where the DNA exhibited scale-invariant breathing, 
and that at even more extreme twist angles, the open state would be 
stable.  However, this is not the case at all, instead, we find $1/f$ 
behaviour at all twist angles.
Since the emergence of this critical behaviour is not affected by the 
variation of the twist angle of the system parameters values, or by the 
careful tuning of other parameters,  we describe this as self-organised 
criticality. 

The scale-free nature of the kinetics of breathing events at all 
twist angles described herein is strongly reminiscent of the 
behaviour of fluctuations in systems at criticality. Thus, it appears 
that {\em without any tuning} of the interaction parameters of
the DNA strand, it is at a critical point where open bubbles 
spontaneously nucleate, hence we apply the term `self-organized 
criticality'.  Figures~\ref{FT_FIG} and \ref{LOGFFT_FCT38_LONG} 
suggest that there is an upper frequency (around $\log_e\omega=-1$), 
above which the amplitude of base-pair separation modes is small 
but ceases to decay any further, due to the effect of noise in the system. 
This cutoff is not due to the start of the phonon band (which 
occupies the range $\sqrt{\gamma} < \omega < \sqrt{\gamma+4k}$), 
and which corresponds to a relatively narrow range of velocities 
around $11<\omega<14$ (precise values depend on the twist angle). 

We observe some artifacts of the phonon band in the region of 
$\log\omega$ being between two and three and the defect mode near 
$\log\omega=1$, in that there is a shoulder in the power spectrum in 
Figure \ref{FT_FIG34} where the trace is slightly larger than expected; 
however, no behaviour should be expected to persist over {\em all} scales, 
and the figures show good agreement with scale-free kinetic behaviour 
(straight-line) over the considerably large range of $-6<\log\omega<1$.  

One might think that the defect site is the cause of the SOC breathing
behaviour. However, replacing the thymine (T) base with a difluorotoluene 
(F) base lowers the barrier between the closed and open states.
The replacement does not affect the DNA structure or other behaviour,
as discussed in several papers, for example, \cite{Guckian}.
Lowering the energy barrier allows breathing to occur at lower energies,
and so occur with a higher frequency, on a timescale accessible to 
MD simulations  (on the nanosecond timescale as opposed to the 
microsecond scale for a normal DNA sequence).  There is no reason 
to suppose that a change in the frequency of events should cause a 
more significant change in qualitative behaviour. Hence, 
we speculate that in pure DNA, with no defect, but with multiwelled
potentials between all corresponding base-pairs, curves such as that 
seen in Figure \ref{FT_FIG}, will be repeated
but that the crossover frequency (from $\omega$-independent noise
to breathing with amplitude proportional to $1/\omega$) will be shifted
to much lower frequencies, namely the microsecond scale, which is
beyond current MD simulations.  Here we only have a 
double-well potential at the defect, the other inter-base 
interactions are all governed by harmonic potentials, in reality, 
all inter-base interactions are double-welled, this will allow 
an open base-pair to be the nucleus for a bubble of several 
consecutive open base-pairs to form, as the along-chain interactions 
would then ease the opening of neighbouring base-pairs. 


\subsection*{Acknowledgments}

CID was funded by the EU as part of MMBNOTT -- an Early Training
Research Programme in Mathematical Medicine and Biology.

\footnotesize

\bibliographystyle{model1-num-names}

\begin{thebibliography}{99}
\parsep=0pt
\itemsep=0pt 

\bibitem{hennig} D.~Hennig - \textit{Formation and propagation
of oscillating bubbles in DNA initiated by structural distortions},
Eur. Phys. J. B 37, 391--397, (2004).

\bibitem{amb} T.~Ambj\"{o}rnsson, S.K.~Banik, M.A.~Lomholt \& 
R.~Metzler - \textit{Master equation approach to DNA breathing in
heteropolymer DNA}, Phys Rev E 75, 021908, (2007).

\bibitem{Amber} D.A.~Case, T.A.~Darden, T.E.~Cheatham, III, 
C.L.~Simmerling, J.~Wang, R.E.~Duke, R.~Luo, K.M.~Merz, D.A.~ 
Pearlman, M.~Crowley, R.C.~Walker, W.~Zhang, B.~Wang, S.~Hayik, 
A.~Roitberg, G.~Seabra, K.F.~Wong, F.~Paesani, X.~Wu, S.~Brozell, 
V.~Tsui, H.~Gohlke, L.~Yang, C.~Tan, J.~Mongan, V.~Hornak, G.~Cui, 
P.~Beroza, D.H.~Mathews, C.~Schafmeister, W.S.~Ross, \& P.A.~Kollman,
\textit{AMBER 9}, University of California, San Francisco (2006).

\bibitem{Duduiala1} C.I.~Duduial\u{a}, J.A.D.~Wattis, I.L.~Dryden \& 
C.A.~Laughton - \textit{Nonlinear breathing modes at a defect site 
in DNA},  Phys Rev E, 80, 061906, (2009).

\bibitem{WattisLaughton} J.A.D.~Wattis, S.A.~Harris, C.R.~Grindon \& 
C.A. Laughton - \textit{Dynamic model of base pair breathing in 
a DNA chain with a defect},  Phys Rev E, 63, 061903, (2001).

\bibitem{Wattis} J.A.D.~Wattis  - \textit{Nonlinear breathing modes 
due to a defect in a DNA chain}, Phil Trans Roy Soc Lond A, 362, 
1461--1477, (2004).

\bibitem{Watson} J.D.~Watson \& F.H.C.~Crick - \textit{Molecular
structure of nucleic acids}, Nature 171, 737 (1953). 

\bibitem{cubero} E.~Cubero, E.C.~Sherer, F.J.~Luque, M.~Orozco \& 
C.A.~Laughton - \textit{Observation of spontaneous base pair 
breathing events in the molecular dynamics simulation of a 
difluorotoluene-containing DNA oligonucleotide}, J. Am. Chem. Soc. 
121, 8653--8654, (1999). 

\bibitem{Peyrard} M.~Peyrard \& A.R.~Bishop - \textit{Statistical
mechanics of a nonlinear model for DNA denaturation},
Phys Rev, 62, 2755-2758 (1989).

\bibitem{PeyrardFarago} M.~Peyrard \& J.~Farago - \textit{Nonlinear
localization in thermalized lattices: Application to DNA},
Physica A 288, 199-217 (2000).

\bibitem{PeyrardLopezJames} M.~Peyrard, S.C.~L\'{o}pez \& G.~James
- \textit{Modelling DNA at the mesoscale: a challenge for nonlinear 
science?},  Nonlinearity 21, T91-T100 (2008).

\bibitem{Gaeta} G.~Gaeta \& L.~Venier - \textit{Solitary waves in
twist-opening models of DNA dynamics}, Phys Rev E 78, 011901 (2008).

\bibitem{Caldarelli} G.~Caldarelli, R.~Frondoni, A.~Gabrielli1,
M.~Montuori1, R.~Retzlaff \& C.~Ricotta - \textit{Percolation in 
real wildfires},  Europhys Lett 56, 510-516 (2001).

\bibitem{Cardoni}  M.~Cadoni, R.~De Leo \& G.~Gaeta - \textit{A 
composite model for DNA torsion dynamics}, Phys Rev E 75 021919 (2007).

\bibitem{Ruelle} D.~Ruelle - \textit{Small random perturbations of
dynamical systems and the definition of attractors},
Commun Math Phys 82, 137-151 (1981).

\bibitem{Wolfram} S.~Wolfram - \textit{Cellular automata as model of
complexity}, Nature 311, 419 (1984).

\bibitem{Haken} H.~Haken - \textit{Cooperative phenomena in
systems far from thermal equilibrium and in nonphysical systems},
Rev Mod Phys 47, 67-121 (1975).

\bibitem{Zinn} J.~Zinn-Justin - \textit{Quantum Field Theory and
Critical Phenomena}, Oxford University Press (2002).

\bibitem{Chopard} B.~Chopard \& M.~Droz - \textit{Cellular Automata
Modeling of Physical Systems}, Cambridge University Press (1998).

\bibitem{Peng} G.~Peng \& D.~Tian - \textit{The fractal nature of a
fracture surface}, J Phys A: Math Gen 23, 3257-3261 (1990).

\bibitem{Newman} M.E.J.~Newman - \textit{Power laws, Pareto
distributions and Zipf's law}, Contemporary Physics 46(5), 323-351 (2005).

\bibitem{Bak87} P.~Bak, C.~Tang \& K.~Wiesenfeld - \textit{ 
Self-organized criticality: An explanation of the 1/f noise},
Phys Rev Lett 59, 381--384, (1987).

\bibitem{Bak88} P.~Bak, C.~Tang \& K.~Wiesenfeld
- \textit{Self-organized criticality}, Phys Rev A 38, 364--374, (1988).

\bibitem{Chapman2} S.C.~Chapman - \textit{Inverse cascade avalanche 
model with limit cycle exhibiting period doubling, intermittency, 
and self-similarity},  Phys Rev E 62, 1905--1911, (2000).

\bibitem{Olami} Z.~Olami, H.J.S.~Feder \& K.~Christensen-\textit{ 
Self-organised criticality in a continuous, nonconservative cellular 
automaton modelling earthquakes}, Phys Rev Lett 68, 1244--1247, (1992). 

\bibitem{BakEarth} P.~Bak, K.~Christensen, L.~Danon \& T.~Scanlon - 
\textit{Unified scaling law for earthquakes}, Phys Rev Lett 88, 178501, (2002).
 
\bibitem{BakFF} P.~Bak, K.~Chen \& C.~Tang - \textit{A forest-fire 
model and some thoughts on turbulence},  Phys Lett A 147, 297--300, (1990).

\bibitem{Drossel} B.~Drossel \& F.~Schwabl - \textit{Self-organized 
critical forest-fire model}, Phys Rev Lett 69, 1629--1632, (1992).

\bibitem{Phillips} J.C.~Phillips - \textit{Scaling and self-organized 
criticality in proteins I}, Proc Natl Acad Sci 106(9), 3107--3112, (2009).

\bibitem{Phillips2} J.C.~Phillips - \textit{Scaling and self-organized 
criticality in proteins II}, Proc Natl Acad Sci 106(9), 3113--3118, (2009).

\bibitem{Werner} G.~Werner - \textit{Metastability, criticality and 
phase transitions in brain and its models}, Biosystems 90, 496--508, (2007).

\bibitem{Selvam2} A.M.~Selvam - \textit{Universal spectrum for DNA 
base C-G frequency distribution in Human chromosomes 1 to 24},
arXiv:physics/0701079 (2007).

\bibitem{Harris} S.A.~Harris, E.~Gavathiotis, M.S.~Searle, M.~Orozco 
\& C.A.~Laughton - \textit{Cooperativity in drug-DNA recognition: 
a molecular dynamics study},  J Am Chem Soc 123, 12658--12663, (2001).

\bibitem{schlitter} J.~Schlitter  - \textit{Estimation of absolute
and relative entropies of macromolecules using the covariance
matrix}, Chem. Phys. Lett. 215, 617--621, (1993).

\bibitem{Duduiala4} C.I.~Duduial\u{a} - \textit{Stochastic Nonlinear 
Models of DNA Breathing at a Defect},  {\tt http://etheses.nottingham.ac.uk/},
PhD Thesis, University of Nottingham, (2009).

\bibitem{PeyrardLopez} M. Peyrard, S. C. L\'{o}pez, D. Angelov 
- \textit{Fluctuations in the DNA double helix}, 
Eur Phys J Special Topics 147, 173-189 (2007).

\bibitem{Guckian} K.M.~Guckian, T.R.~Krugh \& E.T.~Kool - \textit{Solution 
structure of a DNA duplex containing a replicable difluorotoluene-adenine 
pair},  Nature Structural Biology  5, 954--959, (1998).

\end{thebibliography}

\end{document}